 \documentstyle[aas2pp4]{article}
\lefthead{KASPI ET AL.}

\righthead{THE BLR SIZE IN TWO BRIGHT QUASARS}

\def\gtorder{\mathrel{\raise.3ex\hbox{$>$}\mkern-14mu
                \lower0.6ex\hbox{$\sim$}}}
\def\ltorder{\mathrel{\raise.3ex\hbox{$<$}\mkern-14mu
                \lower0.6ex\hbox{$\sim$}}}

\begin{document}

\title{
% . \\
% . \\
% . \\
% . \\
% . \\
% . \\
% MEASUREMENT OF THE BROAD \\ LINE REGION SIZE IN TWO \\ BRIGHT QUASARS}
 MEASUREMENT OF THE BROAD LINE REGION \\ SIZE IN TWO BRIGHT QUASARS}

\author{
Shai~Kaspi,\altaffilmark{1}
\altaffiltext{1}{School of Physics and Astronomy and the Wise
Observatory, The Raymond and Beverly Sackler Faculty of Exact Sciences,
Tel-Aviv University, Tel-Aviv 69978, Israel.}
\authoremail{shai@wise.tau.ac.il}
Paul~S.~Smith,\altaffilmark{2,3}
\altaffiltext{2}{Steward Observatory, University of Arizona, Tucson, AZ 85721.}
\altaffiltext{3}{Current address: NOAO/KPNO, P.O. Box 26732, Tucson, AZ 85726-6732.}
Dan~Maoz,\altaffilmark{1} \\
Hagai~Netzer,\altaffilmark{1}
and Buell.~T.~Jannuzi\altaffilmark{4}
\altaffiltext{4}{NOAO/KPNO, P.O. Box 26732, Tucson, AZ 85726-6732.}
}

\vspace{1cm}

\centerline{Accepted by the $ApJL$}
\vspace{0.5cm}

\centerline{Version of August 19, 1996}
\begin{abstract}

We present 4 years of spectrophotometric monitoring data for two
radio-quiet quasars, PG~0804+762 and PG~0953+414, with typical sampling
intervals of several months. Both sources show continuum and emission
line variations. The variations of the H$\beta$ line follow those of
the continuum with a time lag, as derived from a cross-correlation
analysis, of 93$\pm$30 days for PG~0804+762 and 111$\pm$55 days for
PG~0953+414. This is the first reliable measurement of such a lag in
active galactic nuclei with luminosity $L>10^{45}$~erg~s$^{-1}$. The
broad line region (BLR) size that is implied is almost an order of
magnitude larger than that measured in several Seyfert 1 galaxies and
is consistent with the hypothesis that the BLR size grows as
$L^{0.5}$.

\end{abstract}


\section{Introduction}

Reverberation mapping has became one of the major tools for studying
the distribution and kinematics of the gas in the broad line region
(BLR) of active galactic nuclei (AGN) (see, e.g., Peterson 1993,
Gondhalekar, Horne, \& Peterson 1994). About a dozen Seyfert~1 galaxies
have been successfully monitored so far. The time lags between the
emission line and the continuum light curves measured in these objects
can be interpreted in terms of the delayed response of the
spatially-extended BLR to the ionizing continuum source.  The best
studied Seyfert 1 galaxy, NGC~5548, was monitored from the ground for
over 5 years, and from space for several long periods (Korista et al.
1995, and references therein). Several other Seyfert 1s, such as
NGC~4151 and Mrk~279, were observed for periods of order a year or
less. While these observations have not uniquely determined the
geometries of the BLRs, they have established that Seyfert~1 BLRs have
sizes of the order of a few light-days to light-weeks.

While much progress has been achieved in reverberation mapping of
Seyfert galaxies, relatively little is known about the BLR size in
high-luminosity AGN. Few spectrophotometric monitoring attempts of
quasars have been made and most of these (e.g., Zheng et al 1987; Perez
Penston \& Moles 1989; Gondahalekar 1990; Korista 1991; Jackson et al.
1992) resulted in no clear lag determinations, either due to poor
sampling, or because no line variability was detected. Some monitoring
projects have yielded controversial results, such as, for example, the
{\it International Ultraviolet Explorer} campaign on 3C~273. O'Brien \&
Harris (1991) report a lag of 74 days between the Ly$\alpha$ emission
line and the continuum, while Ulrich et al. (1993) argue that the line
variations are only marginally significant.

Since mid-1991, we have been monitoring a well-defined sub-sample of 28
quasars from the Palomar-Green (PG) sample (Schmidt \& Green 1983) with
typical sampling intervals of $1-4$ months. Results of the first 1.5
years were presented in Maoz et al. (1994; Paper~I) where it was shown
that most quasars had undergone continuum variations in the range of
10\%~--~70\%. Balmer line variations that are correlated with the
continuum changes were detected in several objects.  Based on those
preliminary data, it was demonstrated that the emission-line response
times in several quasars is $\ltorder$\,6 months. Reverberation mapping
of these objects therefore requires several years, with sampling
intervals of less than a few months.  The need for a long temporal
baseline is illustrated by the previous results for PG~0953+414, for
the periods 1987-89 and 1991-92, presented in Paper~I.  Despite
continuum variations of $\sim$35\%, no clear line variations were
detected. As we will show in this {\it Letter}, with a monitoring
campaign of longer duration we detect the line variability in this
object.

We present 4 years of data for two radio-quiet quasars from our sample,
PG~0804+762 and PG~0953+414 (Table~\ref{param}). The two quasars show
clear evidence of a time lag between the Balmer lines and the continuum
variations. We use this to set significant constraints on the BLR size
in these high-luminosity objects. In \S~2 we describe the observations,
present the light curves, and carry out a cross-correlation analysis to
determine the BLR size. In \S~3 we discuss the results.

\begin{deluxetable}{lcc}
\tablecolumns{3}
\tablecaption {OBSERVATIONAL PARAMETERS \label {param}}
\tablewidth{0pt}
\tablehead{
\colhead{} & \colhead{PG~0804+762} & \colhead{PG~0953+414} }
\startdata
$z$         	&  0.100      &  0.239       \nl
$B$ magnitude   & $\sim$15.2  &  $\sim$15.1  \nl
Luminosity\tablenotemark{a}  & $\sim$2$\times$10$^{45}$  
                                     &     $\sim$5$\times$10$^{45}$  \nl
\tablevspace{3mm}
blue continuum band\tablenotemark{b}  & 5224--5264 \AA & 5238--5278 \AA \nl
red continuum band\tablenotemark{b}  & 5598--5630 \AA & 6288--6320 \AA \nl
H$\beta$ range\tablenotemark{b}     & 5266--5474 \AA & 5906--6086 \AA \nl
\enddata
\tablenotetext{a}{Between 0.1--1$\mu$m in units of erg~s$^{-1}$,
assuming a power-law continuum ($f_{\nu}\propto\nu^{-\gamma}$)
normalized at the observed optical flux ($H_0=75$ km
s$^{-1}$Mpc$^{-1}$, $q_0=0.5$, $\gamma=0.5$).}
\tablenotetext{b}{Wavelengths given in observer's frame.}
\end{deluxetable}

\section{Observations and Analysis}
\label{Observations}

The observations and the reduction procedure are described in detail in
Paper~I. We repeat here the main points. The observations were carried
out using the Steward Observatory 2.3m telescope and the Wise
Observatory 1m telescope. For each quasar, the spectrograph slit was
rotated to the appropriate position angle so that a nearby comparison
star was observed along with the object. Wide slits
(4.5$^{\prime\prime}$ at Steward, 10$^{\prime\prime}$ at Wise) were
used to minimize the effects of atmospheric dispersion at the
non-parallactic position angle. The quasar flux is calibrated relative
to that of the comparison star. This technique provides excellent
calibration even during poor weather conditions, and accuracies of
order 1\%~--~2\% can easily be achieved.

Observations typically consisted of two consecutive exposures of the
quasar/star pair. Total exposure times were usually 40 minutes long at
Steward and 2 hours at Wise. The spectroscopic data were reduced using
standard IRAF\footnote{{\tiny {IRAF (Image Reduction and Analysis Facility) is
distributed by the National Optical Astronomy Observatories, which are
operated by AURA, Inc., under cooperative agreement with the National
Science Foundation.}}} routines. The consecutive quasar/star flux ratios
were compared to test for systematic errors in the observations. The
ratios almost always reproduced to 0.5-1.5\% at all wavelengths and
observations with ratios larger than 5\% were discarded.

For both quasars we examined the best S/N spectra and chose line-free
spectral bands suitable for setting the continuum underlying the
emission lines and the wavelength limits for integrating the line
fluxes. The spectral regions for H$\beta$ and the continuum bands are
given in Table~\ref{param}. The line and continuum fluxes were measured
automatically for all epochs by calculating the mean flux in the
continuum bands and summing the flux above a straight line in
$f_{\lambda}$ connecting the continuum bands straddling the emission
line.

\begin{figure}[t]
%\vspace{-0.4cm}
\epsscale{1.00}
\plotone{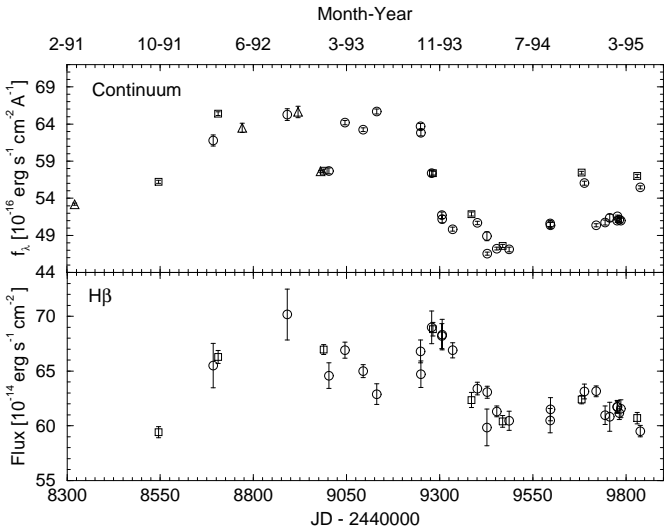}
\vspace{-0.1cm}
\caption{PG~0804+762 light curves. Top panel -- continuum flux density
at 5244 \AA. Bottom panel -- H$\beta$ emission-line flux. Circles are
spectrophotometric measurements from Wise Observatory and squares are
from Steward Observatory.  Triangles are B-band photometric
measurements from Wise Observatory (see Paper~I).}
\label{lc0804}
\end{figure}

Figure~\ref{lc0804} shows the continuum measurements (on the blue side
of H$\beta$) for PG~0804+762, together with $B$-band photometry from
Paper~I, and the H$\beta$ light curve. Observations of this quasar used
different comparison stars at Wise and at Steward (see Paper~I).  Note
the excellent agreement between these completely independent
measurements. The continuum variability (defined as
$[{F_{max}}/{F_{min}}-1]$) is 40\% and the H$\beta$ variability is
$\sim$18\%. The continuum variability time scale, defined as the width
of the continuum auto-correlation function at 0.5 correlation, is 320
days. The H$\beta$ light curve clearly follows the continuum. It is
smoother than the continuum light curve and does not show the shorter
and weaker spikes.  Nearly identical behavior and variability are seen
in the H$\alpha$ light curve (not shown).

\begin{figure}[t]
%\vspace{-0.4cm}
\epsscale{1.00}
\plotone{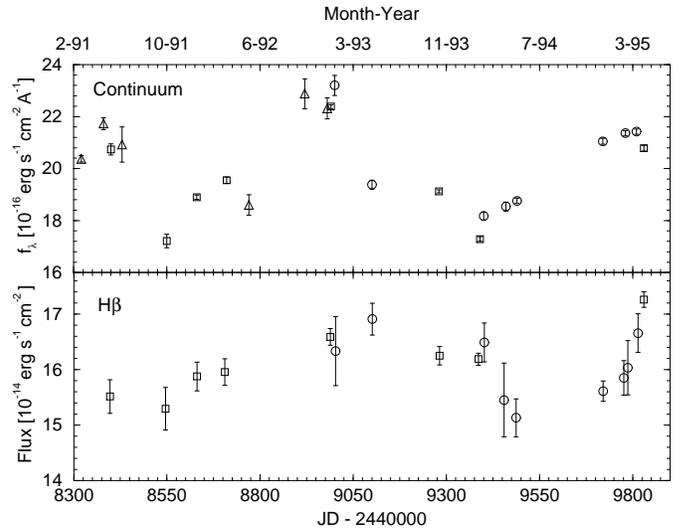}
\vspace{-0.1cm}
\caption{PG~0953+414 light curves. Top panel --
continuum at 5268 \AA.  Bottom panel -- H$\beta$. Symbols as in
Fig.~\protect\ref{lc0804}.} 
\label{lc0953}
\end{figure}

Figure~\ref{lc0953} \ shows the \ corresponding \ light curves \ for
PG~0953+414.  During the 4-year period, the continuum varied by 35\%
while the H$\beta$ light curve, which appears to follow the continuum,
has a typical amplitude of 13\%. The continuum variability time scale,
defined as above, is 200 days. The H$\gamma$ line (not shown here)
exhibits similar variations to H$\beta$.

\begin{deluxetable}{lcc}
\tablecolumns{3}
\tablecaption {CROSS CORRELATION FUNCTION TIME LAGS (DAYS) \label {peaks}}
\tablewidth{0pt}
\tablehead{
\colhead{} & \colhead{PG~0804+762} & \colhead{PG~0953+414} }
\startdata
PICCF\tablenotemark{a} \ centroid     &	102  &  115  \nl
ZDCF\tablenotemark{a}  \ centroid     &	84   &  107  \nl
BLR size\tablenotemark{b}             &   85$\pm$27   & 90$\pm$44    \nl
\enddata
\tablenotetext{a}{See \S~2 for definition.}
\tablenotetext{b}{Deduced from averaging centroid results and applying
a $(1+z)^{-1}$ factor.}
\end{deluxetable}

\begin{figure}[t]
\vspace{0.4cm}
\epsscale{1.40}
\plotone{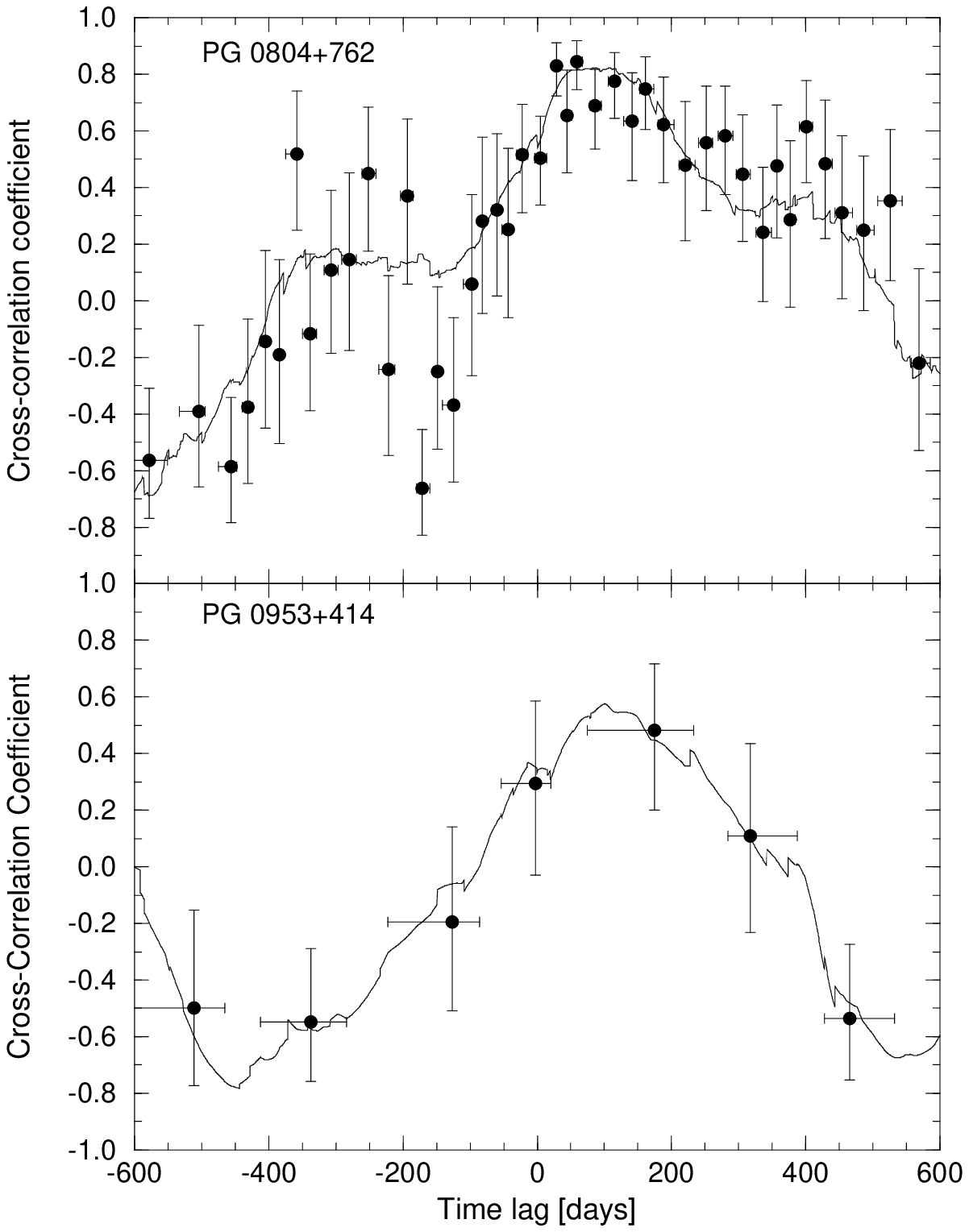}
\vspace{-0.3cm}
\caption{PICCF (solid line) and ZDCF (circles with error bars) of the
H$\beta$ light curve with the continuum light curve for PG~0804+762
(top) and for PG~0953+414 (bottom).  Horizontal error bars indicate
1$\sigma$ range of the ZDCF bin size (i.e., the standard deviation of
time differences included in the bin).}
\label{ccf}
\end{figure}

We have used two methods for correlating the line and continuum light
curves. The partly interpolated cross-correlation function (PICCF) of
Gaskell \& Peterson (1987) and Gaskell (1994) and the $z$-transform
discrete correlation function (ZDCF) of Alexander (1996). The second
method, which is an improvement of the discrete correlation function
(DCF; Edelson \& Krolik 1988), applies Fisher's $z$ transformation to
the correlation coefficients, and uses equal population bins rather
than the equal time bins used in the DCF.  The cross correlations of
the H$\beta$ and the continuum light curves for the two objects are
shown in Figure~\ref{ccf}. Both cross-correlation methods yield similar
results. The ZDCF for PG~0953+414 has few points since its light curves
contain fewer measurements and the number of bins scales as the square
of the number of measurements. Nevertheless, its agreement with the
PICCF is good and indicates that the interpolation done in the PICCF
did not introduce an artificial correlation.  In both quasars, the
H$\beta$ flux lags the continuum by a few months and the
cross-correlation function (CCF) peak is highly significant.  We list
in Table~\ref{peaks} the centroid of all points above 60\% of the peak
correlation. Since both correlation methods yield similar centroid
results, we adopt the mean lag implied by the two methods as our
measure of the BLR radius. We defer to a future paper discussion of
various complications involved in associating an observed time-lag with
a BLR size, e.g., the dependence of the peak of the CCF on the nature
of the continuum variability, the possible non-linear response of the
Balmer line intensity to the continuum flux variations, or the
possibility that the ionizing continuum may behave differently from the
observed continuum.

To estimate the uncertainty in the PICCF time lags, we have carried out
Monte-Carlo simulations as described in Maoz \& Netzer (1989). For each
object the simulation involves a linear interpolation of the observed
continuum light curve and the calculation of the expected line
light-curve for a chosen BLR geometry. These continuum and line light
curves are then repeatedly sampled at random in a seasonal pattern
resembling the observing sequence, simulated measurement errors are
added, and the PICCF and its centroid location are calculated for each
simulated pair. We have tried a variety of spherical geometries and
computed, for each, the cross-correlation centroid distribution.  The
distribution is approximately Gaussian.  The width of the distribution
is a measure of the uncertainty in the lag.  The central width that
contains 68\% of all expected lags is an estimate of the 1$\sigma$
error. This corresponds to $\pm$30 days for PG~0804+762 and $\pm$55
days for PG~0953+414. These uncertainties are adopted in the discussion
below.

\section{Discussion}

Spectrophotometric monitoring of PG~0804+762 and PG~0953+414 have
revealed a clear correlation and lag between the continuum and
Balmer-line light curves.  All Seyfert 1 galaxies that have been
studied in this way have time-averaged luminosities of
4$\times$10$^{44}$~erg~s$^{-1}$ at most, while the PG quasars yield
measures of the time lag in AGN with luminosities exceeding
$10^{45}$~erg~s$^{-1}$. We have found that the H$\beta$ emission line
lags the continuum by 93$\pm$30 days in PG~0804+762 and by 111$\pm$55
days in PG~0953+414. Since measured AGN time lags are an indicator of
the characteristic BLR size, our results allow us to search for a
correlation of the BLR size with source luminosity.  When comparing
lags, the same emission lines should be used in all objects since
different emission lines can have different time lags in a given object
(see, e.g., Peterson 1993). Balmer lines are useful since their lag
has been measured in most AGN where reverberation mapping has been
attempted.

\begin{figure}[t]
\vspace{-0.cm}
\epsscale{1.13}
\plotone{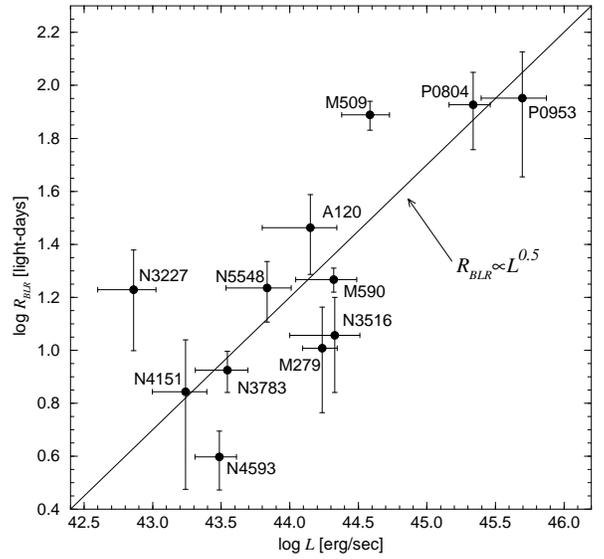}
\vspace{-0.4cm}
\caption{The BLR radius -- luminosity relation. All sizes based on
Balmer-line lags. References: NGC~5548 - Korista et al.  (1995) and
references therein; NGC~4151 - Maoz et al. (1991) and Kaspi et al.
(1996); NGC~3227 -- Salamanca, et al. (1994); NGC~3783 -- Stirpe et
al.  (1994a); Akn~120 -- Peterson (1988) and Peterson \& Gaskell
(1991); Mrk~279 -- Maoz et al. (1990) and Stirpe et al.  (1994b);
NGC~3516 -- Wanders et al. (1993); Mrk~590 -- Peterson et al.  (1993);
NGC4593 -- Dietrich et al. (1994); Mrk~509 -- Carone et al. (1996);
PG~0804+762 and PG~0953+414 -- this work.}
\label{rvsl}
\end{figure}

Figure~\ref{rvsl} compares the BLR size of 12 AGN, as deduced from
cross-correlating the Balmer lines with the optical continuum light
curves, to their 0.1--1~$\mu$m luminosity (defined in
Table~\ref{param}). We restrict ourselves to measurements where a
significant correlation of the line and continuum light curves has been
detected. The BLR radius, $R_{BLR}$, is the mean of values in the
literature for the time lag of a given object, corrected by a
$(1+z)^{-1}$ factor.  Caution must be taken when interpreting the
diagram, since each study has used its own method to deduce $R_{BLR}$
and its uncertainty. The error bars on $R_{BLR}$ are a combination of
the quoted uncertainties in the various references, as well as the
spread in values reported in the literature for a given object. The
uncertainty in the luminosity, $L$, is set by the observed variability
range. For the Seyfert nuclei ($\log L<44.5$) alone there is no clear
correlation given the narrow luminosity range sampled. Adding the
results for PG~0804+762 and PG~0953+414 introduces a possible trend,
and indicates that $R_{BLR}\/$ may scale with $L\/$.  Under the simple
assumptions that the shape of the ionizing continuum in AGN is
independent of $L$, and that all AGN are characterized by the same
ionization parameter and BLR gas density (as indicated by the
generally-similar observed line ratios), $R_{BLR}\propto L^{0.5}$ is
expected. A line with this slope is shown in Figure~\ref{rvsl}. While
this is not the result of a proper fit to the data, such a trend is
consistent with our results. The relation between $R_{BLR}\/$ and $L\/$
(if one exists), yet to be determined from our complete quasar sample,
may hold important clues to the nature of these objects. In particular,
using line-profile information, an AGN mass-luminosity relation can
eventually be derived.

% \vspace{-0.2cm}

\acknowledgments 

We thank John Dan and the Wise Observatory staff for their expert
assistance with the observations. We are grateful to Tal Alexander for
illuminating discussions and the anonymous referee for several
constructive comments.  Astronomy at the Wise Observatory is supported
by grants from the Israel Academy of Science. Monitoring of PG quasars
at Steward Observatory is supported by NASA grant NAG 5-1630.
B.\,T.\,J.  acknowledges support from the National Optical Astronomy
Observatories which is operated by AURA, Inc., on behalf of the
National Science Foundation.

\end{document}